\documentclass[12pt]{article}
\usepackage[dvips]{color}
\usepackage{epsfig}
\usepackage{amsmath}
\usepackage{graphicx}
\begin{document}
\title{\bf The  solution of tachyon inflation in  curved universe}

\author{J.  Sadeghi $^{a}$\thanks{Email:pouriya@ipm.ir}\hspace{1mm}, A. R.  Amani $^{b}
$\thanks{Email:a.r.amani@iauamol.ac.ir} \\
$^a$ {\small {\em Sciences Faculty, Department of Physics, Mazandaran University,}}\\
         {\small {\em P.O.Box 47415-416, Babolsar, Iran}}\\
         $^b$ {\small {\em  Department of Physics, Islamic Azad University - Ayatollah Amoli Branch,}}\\
        {\small {\em P.O.Box 678, Amol, Iran}}}

\maketitle

\begin{abstract}
\noindent In this paper, we have  considered the  curved universe
which  is filled by tachyonic field. We have found the exact
solutions for the field, pressure, density, and scale factor and
some cosmological parameters. In such universe, we have investigated
the role of tachyonic field in different stages of $k$ for the
evolution of the universe.  Finally we draw the graphs for the scale
factor, Hubble's parameter,  energy density, pressure, acceleration
parameter, equation of state and potential for the different values
of  $k.$ Also we obtained  the exact form of field which shows that
the tachyonic field has the kink
form.\\\\

 \end{abstract}

\section{Introduction}

In recent decade, it is observed that the space of the universe
increases with an acceleration rate. Expansion of the universe arise
from energy  called  dark energy, and it is almost three-quarters of
the total mass-energy of the universe. As we know there are few more
taxing questions facing cosmology today than what is nature of the
dark energy in the universe?. Over the past few years there have
been many papers devoted in addressing the nature of the dark energy
and accelerated expansion of universe. On the other hand there have
been difficulties in obtaining accelerated expansion from
fundamental theories such as string theory [1]. Much has been
written and emphasized about the role of the fundamental dilation
field in the context of string
cosmology, but not much emphasis on tachyon component.\\
In this paper, we turn our attention to the issue of the tachyon as
a source of the dark energy. As we know the tachyon is an unstable
field which has become important in string theory through its role
in the Dirac - Born- Infeld (DBI) action which is used to describe
the $D$- brane action [2-4]. A number of authors have already
demonstrated that the tachyon could play an important role in
cosmology [5], independent of the fact that it is an unstable field.
It can act as a source of dark matter and can lead to a period of
inflation depending on the form of the associated potential. Indeed
it has been proposed as the source of dark energy for a particular
class of potential [6-8]. However, there has not really been an
effect to understand the general properties of tachyonic
cosmologies. In order to attempt the problem, we start four-
dimensional DBI action where the tachyon field is
coupled to a background of perfect fluid with  radiation or matter.\\
Also, we consider model of Friedmann-Robertson-Walker (FRW)
cosmology with curvature, driven by real scalar field which evolves
with standard or tachyonic dynamics. We note that in  the Ref. [9],
they have assumed that the universe is filled in tachyonic field
with potential and discussed the acceleration of the universe. In
that case they have considered the FRW model with metric as a
spatially flat. We have assumed that the universe is filled in only
tachyonic field but the FRW metric is curved. So, for tachyonic
dynamics we extend the result of [9], which was in flat space -
time. In contrast to the  paper assuming the tachyon potential,   we
have also obtained the explicit form of potential for the tachyon
field. This paper is organized as follows: section 2 we study
tachyon dynamic with FRW standard model in curved universe . With
the help of energy - momentum tensor we obtain the equation of
motion for the corresponding tachyonic field. The solution for the
tachyonic equation is presented in section 3. In this section we
obtain the exact solution for some cosmological parameters in curved
universe and show  the accelerating expansion of our universe due to
tachyonic field. Finally, in section 4 offer some closing remarks
and results.
\section{Tachyon Dynamic}
One of considerable methods inflationary describe potential or
vacuum energy of scalar field with tachyon field. We can see details
of tachyon field in Refs. [10-12]. However, we want to investigate a
cosmological scenario to help tachyon field $T$ in a perfect fluid.
Now we begin the single tachyonic inflation model with applying
Lagrange density in the DBI type action [13-15]. So, the action for
the homogenous tachyon condensate of string theory in gravitational
background is given by,

\begin{equation}\label{T1}
S=\int d^4x \sqrt{-g} \left (\frac{M^2_p}{2}R +\mathcal{L}\right),
\end{equation}
where $M_p=(8\pi G)^{-1/2}$ is reduced Plank's mass when we  take
$M_p=1/\sqrt{2}$. Also $R$ and $\mathcal{L}$ are scalar
curvature and Lagrangian density respectively.\\ The  lagrangian density   $\mathcal{L}$ is ,\\
\begin{equation}
\mathcal{L}=-
 V(T)\sqrt{1-\partial_{\mu}T\partial^{\mu}T},
\end{equation}
where $V(T)$ is  tachyonic potential. We use spatially metric in FRW
standard model as,
\begin{equation}\label{T1}
ds^2=-dt^2+a^2(t)\left (\frac{dr^2}{1-k~r^2}+r^2~d\Omega^2\right),
\end{equation}
where constant of $k=1,0$ and $-1$ is for spherical, flat and
hyperbolic geometry respectively. \\
We obtain Ricci tensor,
$R_{\mu \nu}$, and Ricci scalar, $R$, as functional of scale factor,
$a(t)$, by,
\begin{equation}
R_{00}=-3\frac{\ddot{a}}{a},
\end{equation}
\begin{equation}
R_{ij}=g_{ij}(\frac{\ddot{a}}{a}+2\frac{\dot{a}^2}{a^2}+2\frac{k}{a^2}),~~~~~i,j=1,2,3,
\end{equation}

\begin{equation}
\alpha^2=2(\frac{\ddot{a}}{a}+\frac{\dot{a}^2}{a^2}+\frac{k}{a^2}),~~~that~~~\alpha=\sqrt{R/3}.\\
\end{equation}
The energy momentum tensor for the tachonic field is,

\begin{equation}
T_{\mu\nu}= -\frac{2}{\sqrt{-g}}\frac{\delta
S}{\delta{g^{\mu\nu}}}=-V(T)\sqrt{1+g^{\mu\nu}\partial_{\mu}T\partial^{\nu
}T}g^{\mu\nu}+\frac{V(T)\partial_{\mu}T\partial_{\nu}T}
{\sqrt{1+g^{\mu\nu}\partial_{\mu}T\partial^{\nu}T}}.
\end{equation}
In FRW metric with a scale factor $a$, the pressure and energy
densities of the field $T$ are given respectively,
\begin{equation}
\rho=\frac{V(T)}{\sqrt{1-\dot{T}^2}},~~~~~
p=-V(T)~\sqrt{1-\dot{T}^2}.
\end{equation}
In order to obtain the corresponding tachyonic field equation we use
energy-momentum tensor as $T^\mu_\nu=(-\rho,p,p,p)$ and Einstein's
equations as $G^\mu _\nu=2~T^\mu_ \nu$, so we have,
\begin{equation}
H^2=\left(\frac{\dot{a}}{a}\right)^2=\frac{2}{3}\rho-\frac{k}{a^2}=\frac{2}{3}\frac{V}{\sqrt{1-\dot{T}^2}}-\frac{k}{a^2},
\end{equation}
\begin{equation}
\dot{H}=-(p+\rho)+\frac{k}{a^2}=-\frac{V~\dot{T}^2}{\sqrt{1-\dot{T}^2}}+\frac{k}{a^2},
\end{equation}
where $ H $ is Hubble's parameter.\\  We rewrite  the equation of
motion as,
\begin{equation}
\frac{\ddot{T}}{1-\dot{T}^2}+3H\dot{T}+\frac{1}{V}\frac{dV}{dT}=0.\\
\end{equation}
We note that the tachyonic potential  and field in terms of scale
factor can be written by  the following respectively,
\begin{equation}
V=\frac{\sqrt{3}}{2}~\sqrt{\frac{\alpha^4}{4}-\frac{\ddot{a}^2}{a^2}},
\end{equation}
\begin{equation}
\dot{T}^2=\frac{2}{3}~\left(1+\frac{1}{1-\frac{\alpha^2}{2}\frac{a}{\ddot{a}}}\right).\\
\end{equation}
As we know, the acceleration parameter and equation state are
respectively ,
\begin{equation}
q=-\frac{a\ddot{a}}{\dot{a}^2}=-1-\frac{\dot{H}}{H^2}.
\end{equation}

\begin{equation}
\omega=\frac{p}{\rho}.
\end{equation}
so we have,
\begin{equation}
\omega=\dot{T}^2-1,
\end{equation}
$\omega$ is a number depending on fluid types, in that case
$\omega=0$  for dust and   $\omega=1/3$  for radiation. Also
condition of universe expansion is $\omega<-1/3$, as it express an
unknown energy that is called dark energy.
\section{  Solution for  the Curved Universe}
By using equation (6) and choosing $\alpha=Constant$, the scale
factor can be written by,
\begin{equation}
a(t)=\pm \frac{1}{\alpha}~\sqrt{2k-\alpha (c_1 e^{\alpha t}-c_2
e^{-\alpha t})},
\end{equation}
where $k$ is curved geometry, $c_1, c_2$ are integration constants
and
$\alpha$ is functional of Ricci scalar $(R)$. For simplicity, here we take $c_2=-c_1=c$ and rewrite scale factor as,   \\
\begin{equation}
a(t)=\pm \frac{\sqrt{2}}{\alpha}~\sqrt{k+\alpha c \cosh(\alpha t)}.
\end{equation}
Now we draw the variation of scale factor with respect to time and
we have fig. (1). It show that scale factor increase as a positive,
so the universe expands eternally.

\begin{tabular*}{2cm}{cc}
\hspace{0.25cm}\includegraphics[scale=0.25]{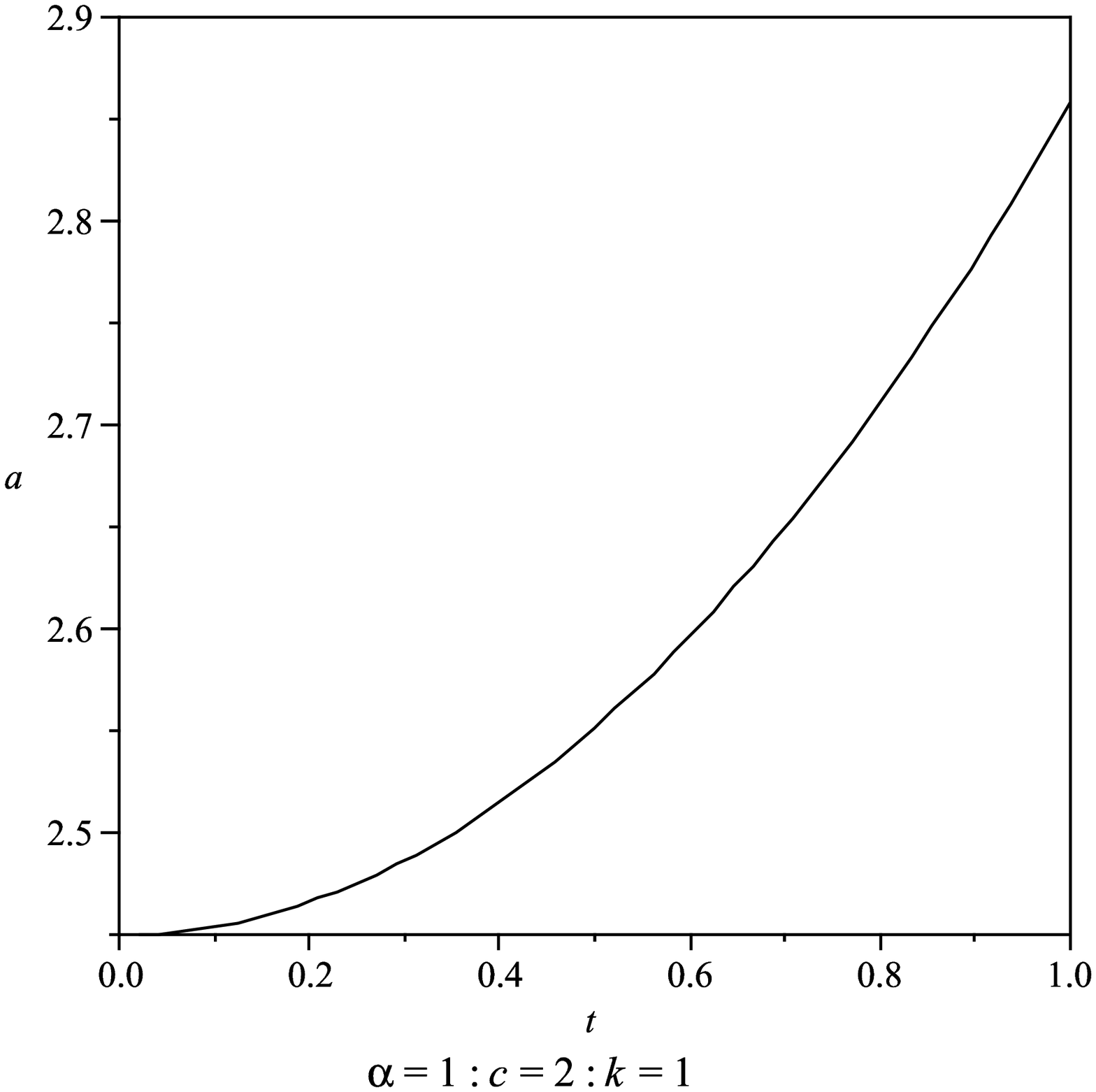}\hspace{0.5cm}\includegraphics[scale=0.25]{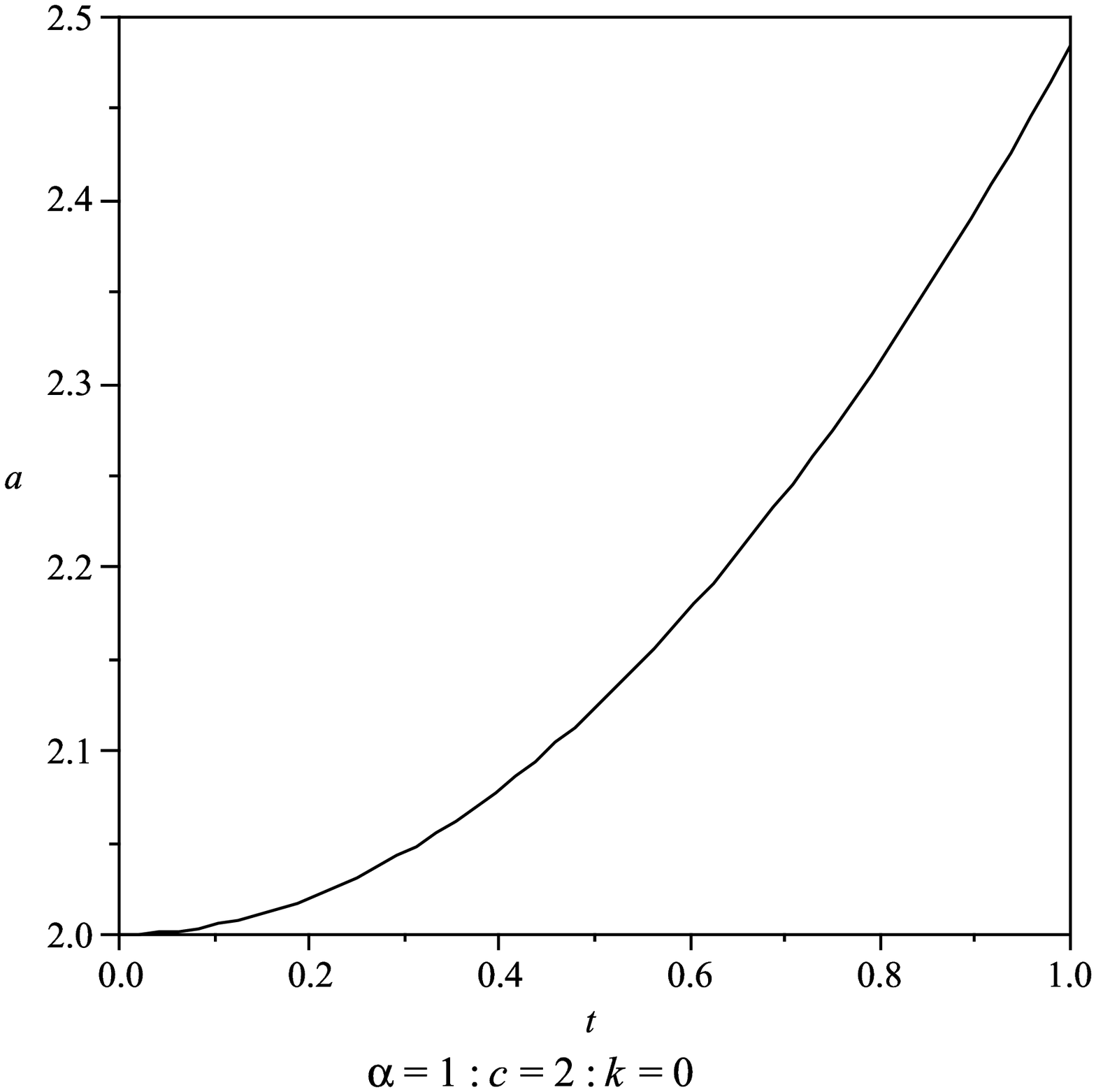}\hspace{0.5cm}
\includegraphics[scale=0.25]{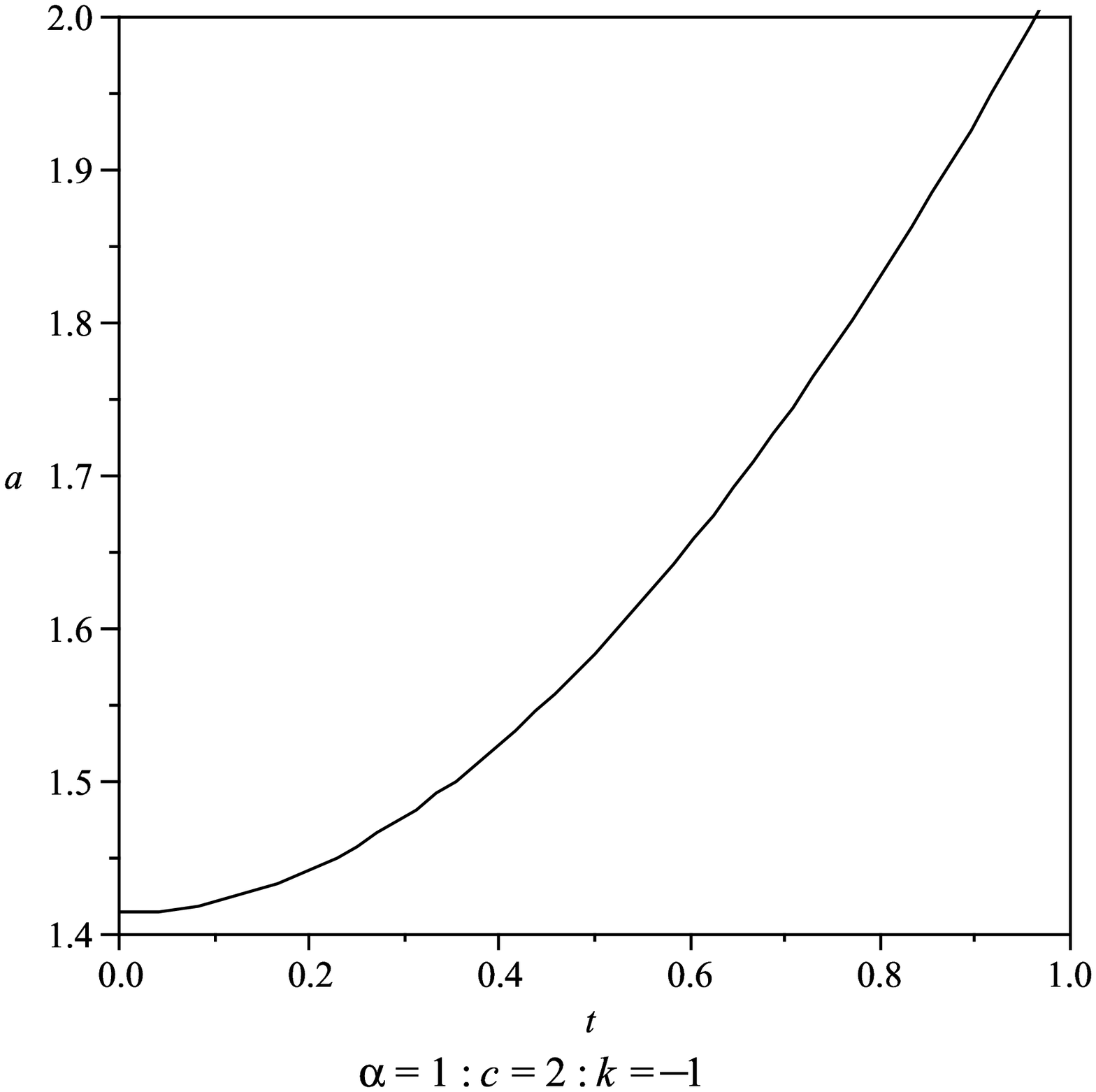}\\
\hspace{-1.5cm}\textbf{Figure 1:} \,Graphs of the scale
factor. \\
\end{tabular*}\\\\\\
The Hubble's parameter is obtained by equation (9) as follows,
\begin{equation}
H(t)=\frac{1}{2}~\frac{\alpha^2 c \sinh(\alpha t)}{k+\alpha c
\cosh(\alpha t)}.
\end{equation}\\
Also we draw the Hubble's parameter in term of time, see fig. (2).
The Hubble's parameter with the choice of different values of $k$, asymptotically increase with constant value.\\

\begin{tabular*}{2cm}{cc}
\hspace{0.25cm}\includegraphics[scale=0.25]{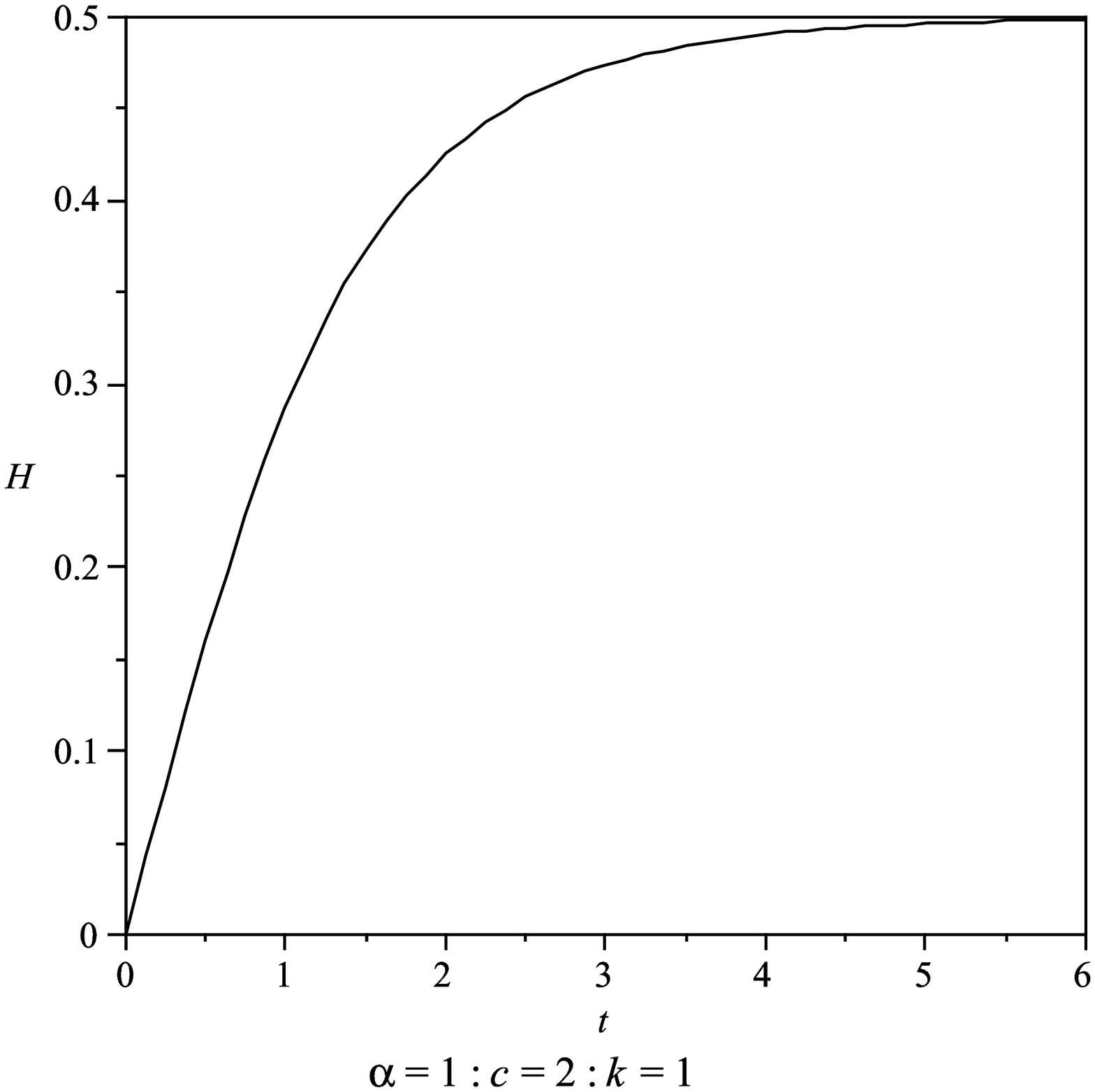}\hspace{0.5cm}\includegraphics[scale=0.25]{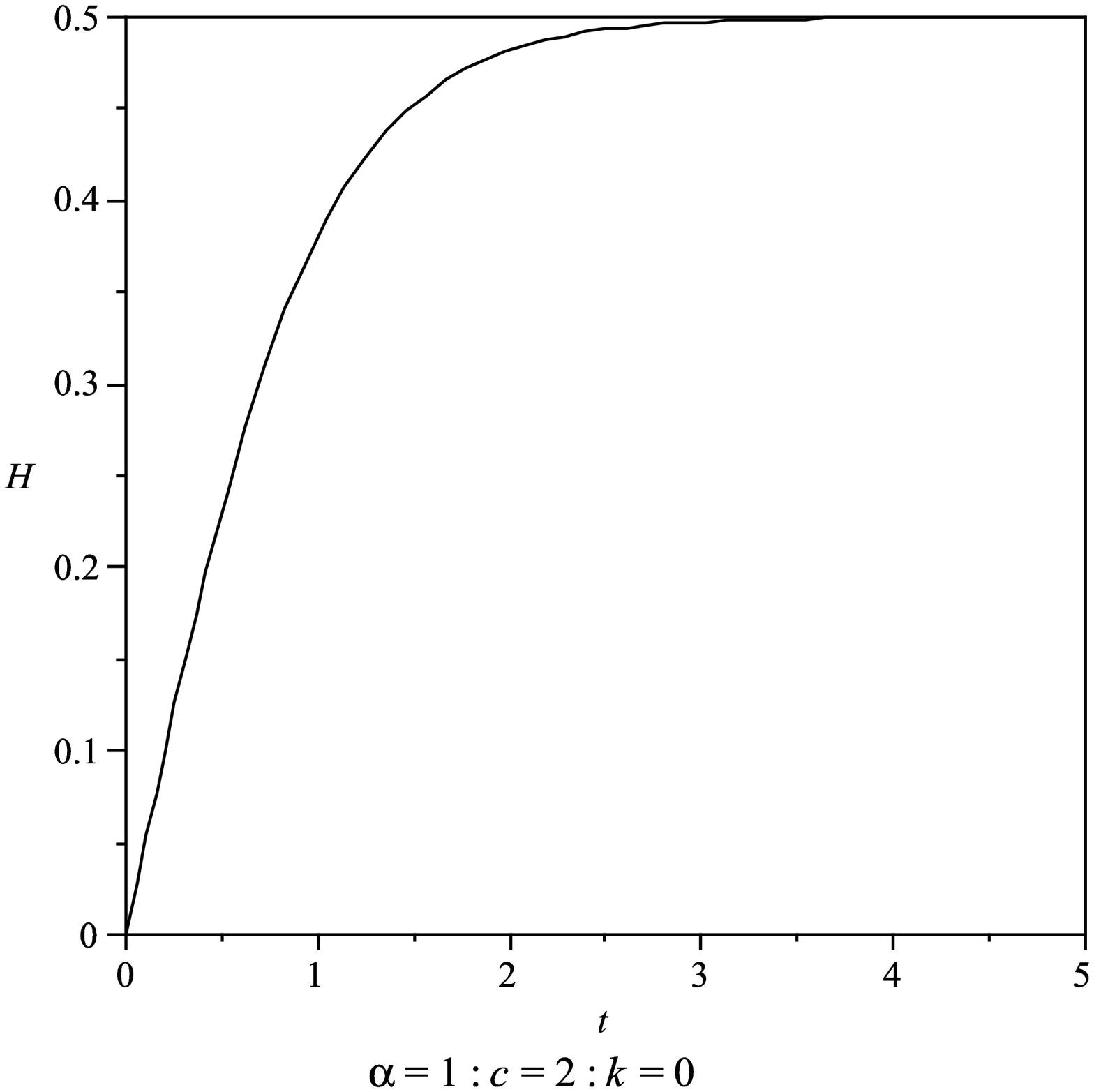}\hspace{0.5cm}
\includegraphics[scale=0.25]{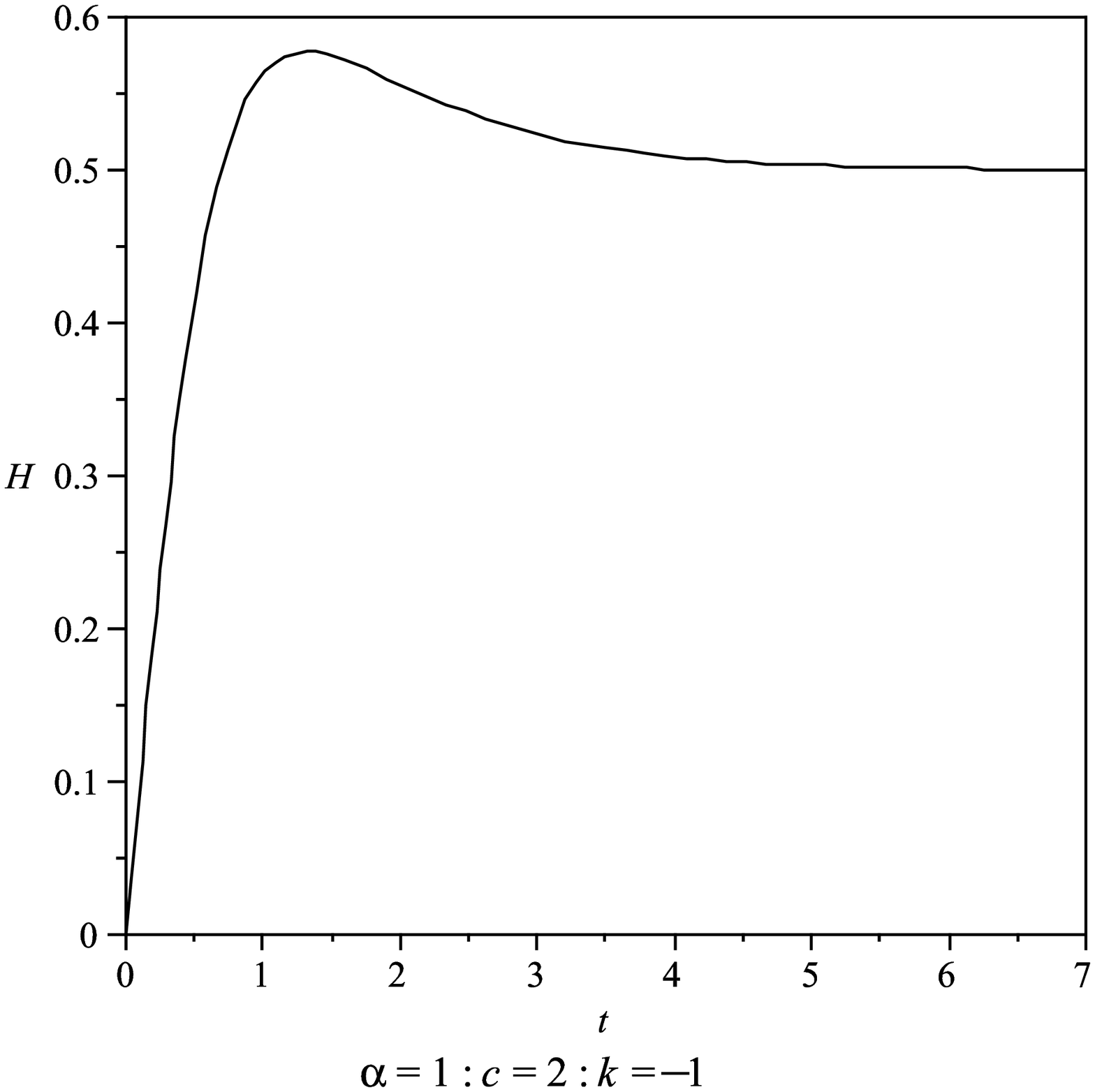}\\
\hspace{-1.5cm}\textbf{Figure 2:} \,Graphs of the Hubble's parameter.\\
\end{tabular*}\\\\\\
The energy density and pressure are obtained as,
\begin{equation}
\rho(t)=\frac{3\alpha^2}{8}~\frac{\alpha^2 c^2 \cosh^2(\alpha
t)+2k\alpha c \cosh(\alpha t)+2k^2-\alpha^2 c^2}{(k+\alpha c
\cosh(\alpha t))^2}.
\end{equation}
\begin{equation}
p(t)=-\frac{\alpha^2}{8}~\frac{3\alpha^2 c^2 \cosh^2(\alpha
t)+6k\alpha c \cosh(\alpha t)+2k^2+\alpha^2 c^2}{(k+\alpha c
\cosh(\alpha t))^2}.
\end{equation}\\
The variation of energy density with respect to time is plotted in
fig. (3). The energy densities for $k=0$ and $k=1$ start to increase
from positive value and tend to asymptotically increase to a value
of
positive constant.\\
\begin{tabular*}{2cm}{cc}
\hspace{0.25cm}\includegraphics[scale=0.25]{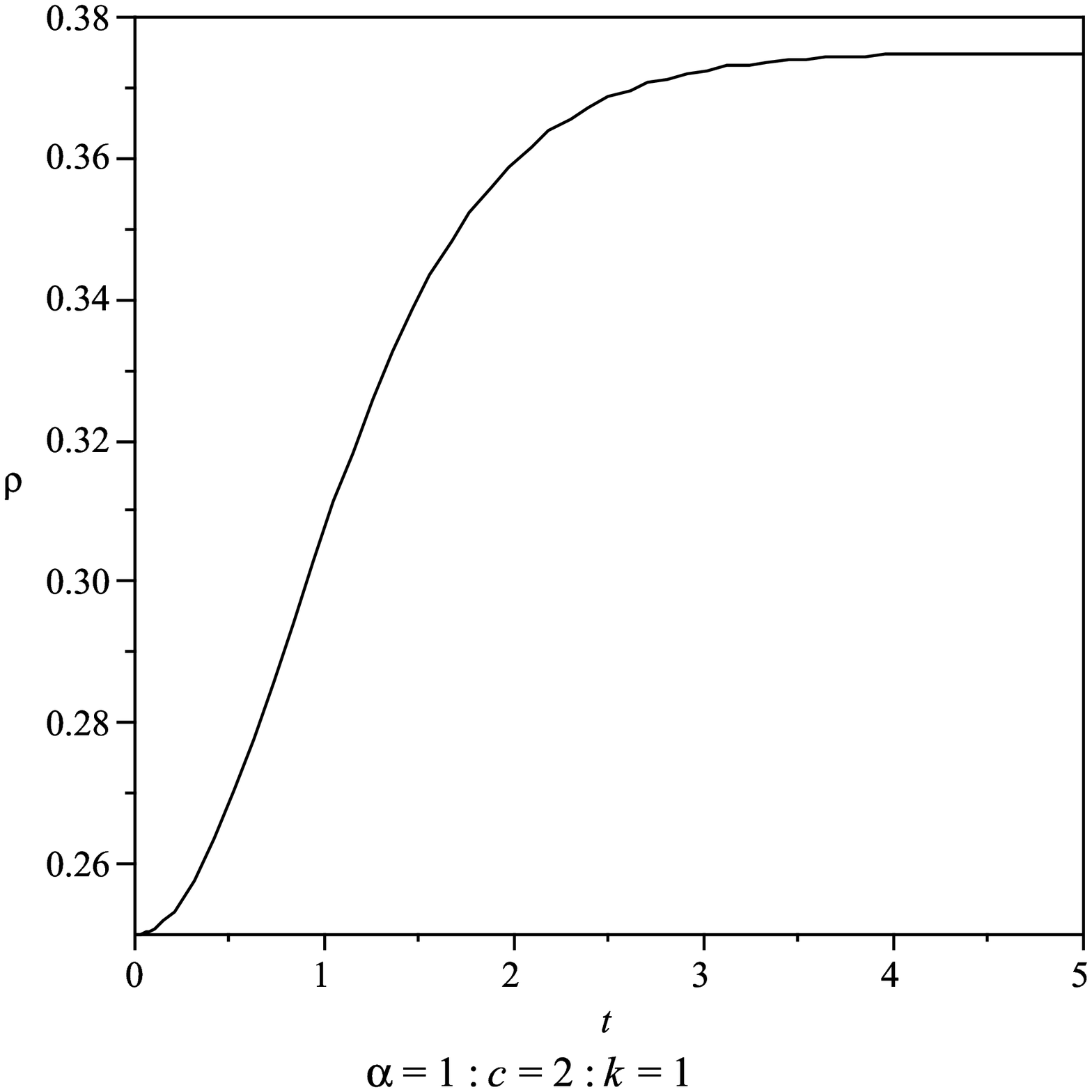}\hspace{0.5cm}\includegraphics[scale=0.25]{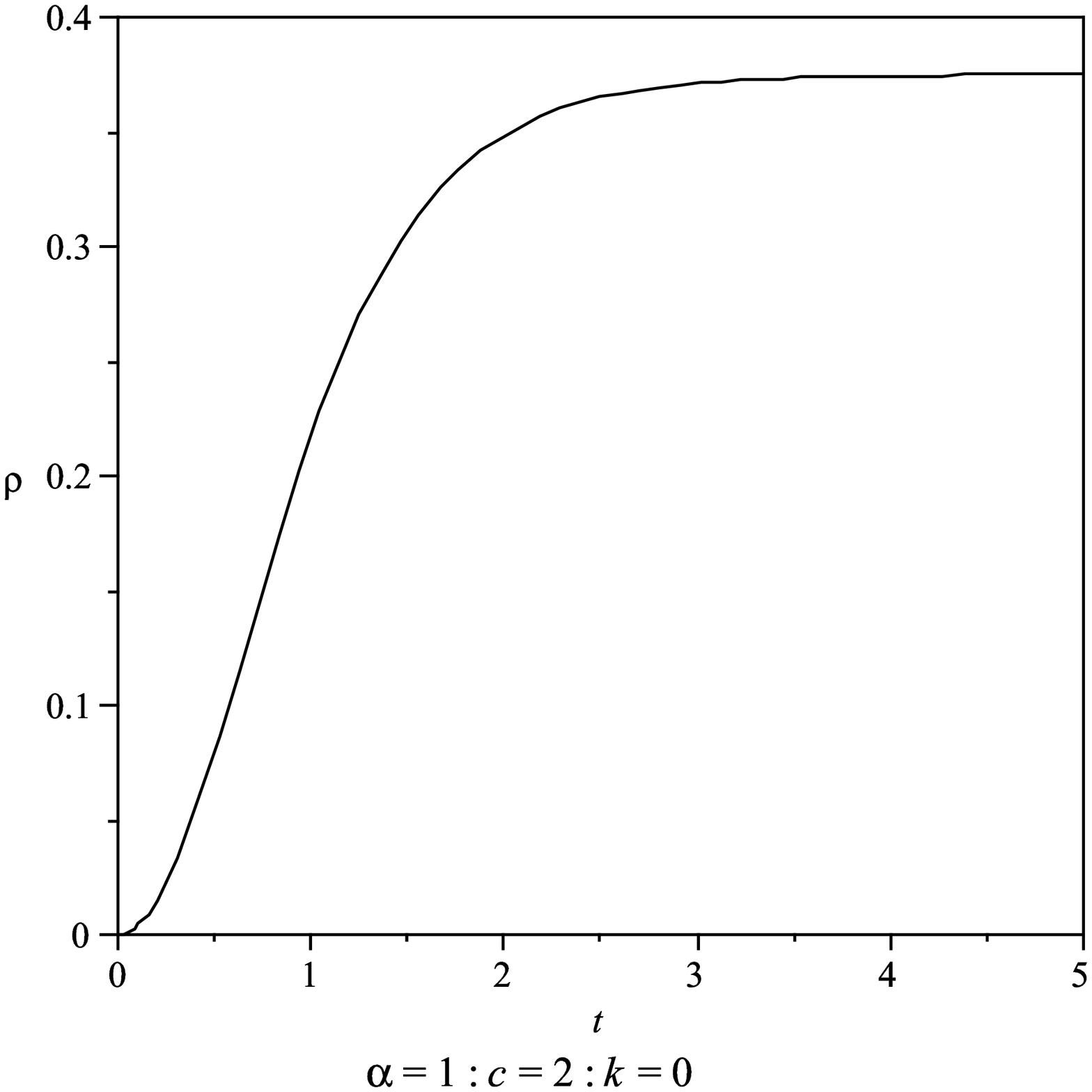}\hspace{0.5cm}
\includegraphics[scale=0.25]{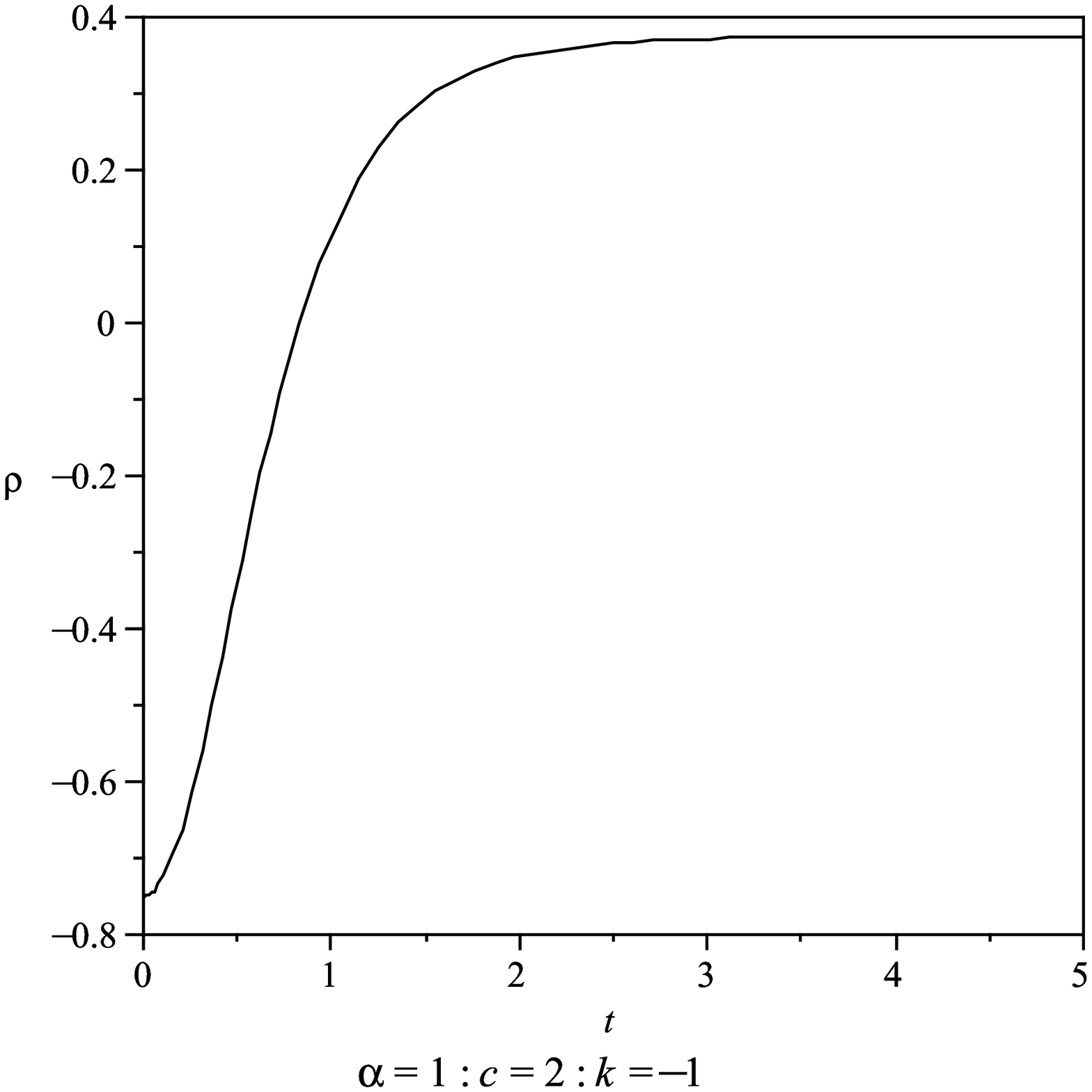}\\
\hspace{-1.5cm}\textbf{Figure 3:} \,Graphs of the energy density.\\
\end{tabular*}\\\\\\
Also we note that in fig. (4) for the different value of $k$ the
pressure start to increase asymptotically to negative value.\\
\begin{tabular*}{2cm}{cc}
\hspace{0.25cm}\includegraphics[scale=0.25]{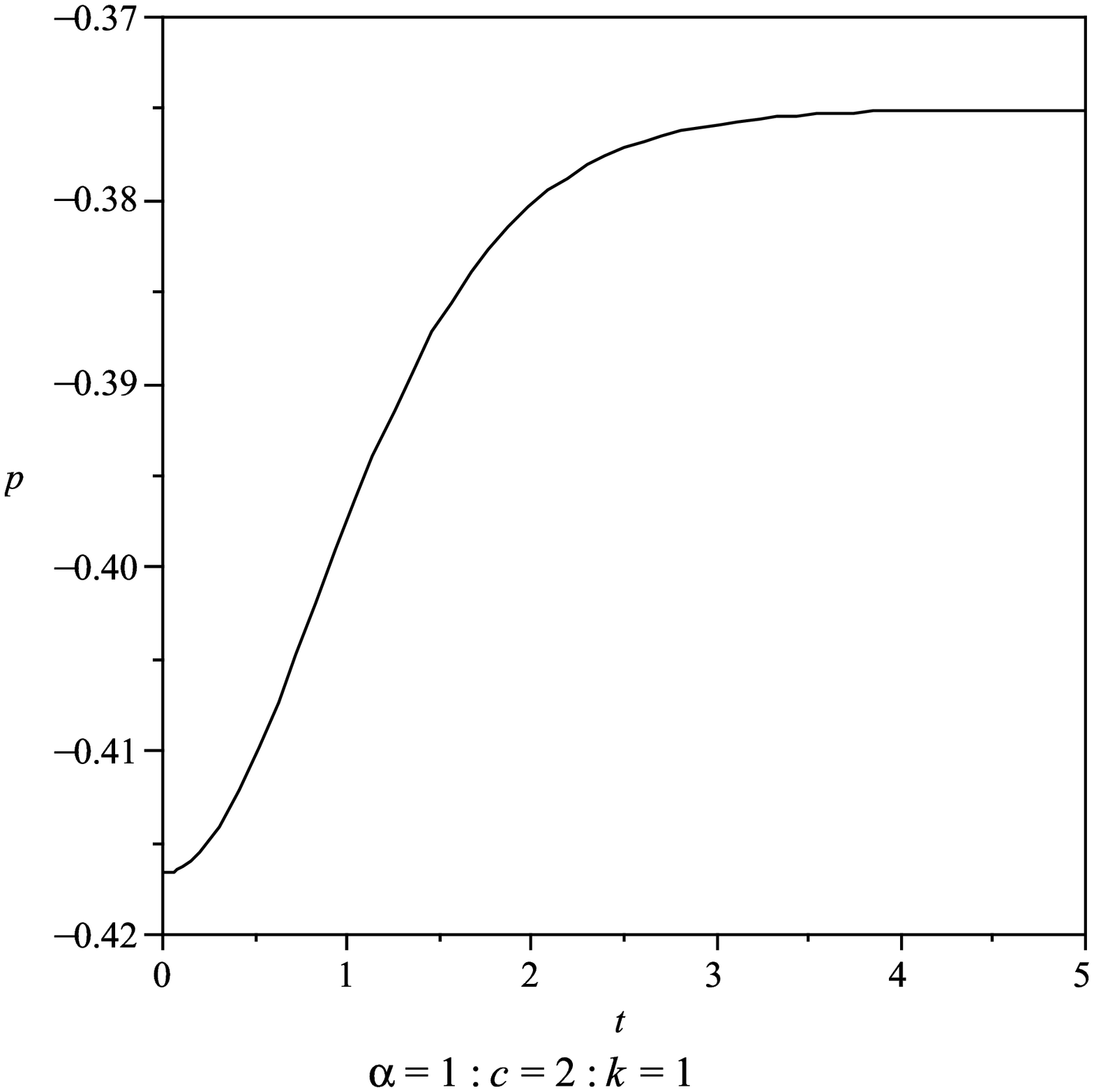}\hspace{0.5cm}\includegraphics[scale=0.25]{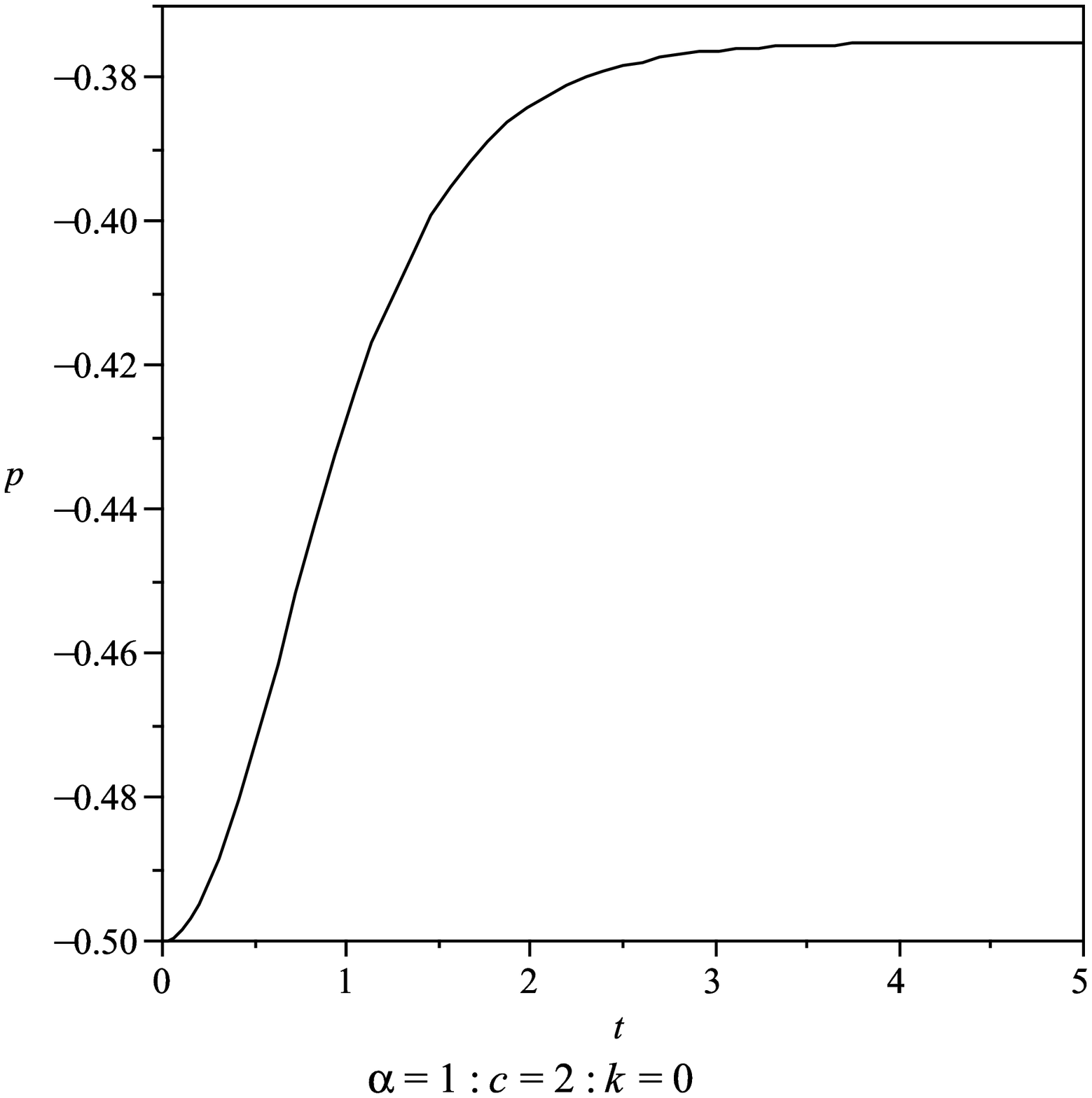}\hspace{0.5cm}
\includegraphics[scale=0.25]{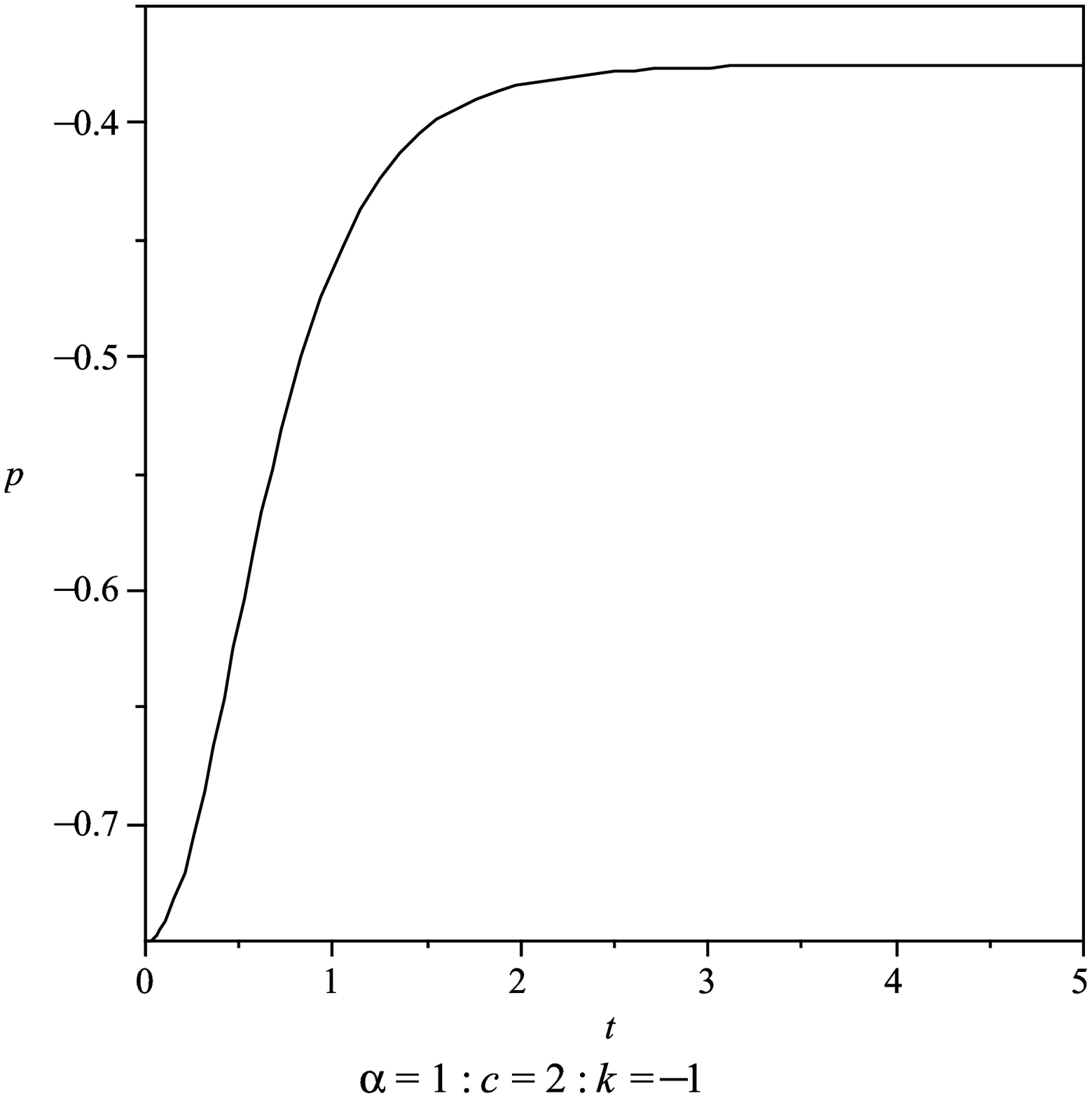}\\
\hspace{-1.5cm}\textbf{Figure 4:} \,Graphs of the pressure.\\
\end{tabular*}\\\\\\
From equations (14) and (18) the acceleration parameter can be
written as,
\begin{equation}
q=-\frac{\ddot{a}a}{\dot{a}^2}=-\frac{c\alpha \cosh^2(\alpha t)+2k
\cosh(\alpha t)+c\alpha}{c\alpha \sinh^2(\alpha t)}.
\end{equation}\\
In fig.(5), we see acceleration parameters increase from a negative
value and tend to asymptotically to $-1$.\\
\begin{tabular*}{2cm}{cc}
\hspace{0.25cm}\includegraphics[scale=0.25]{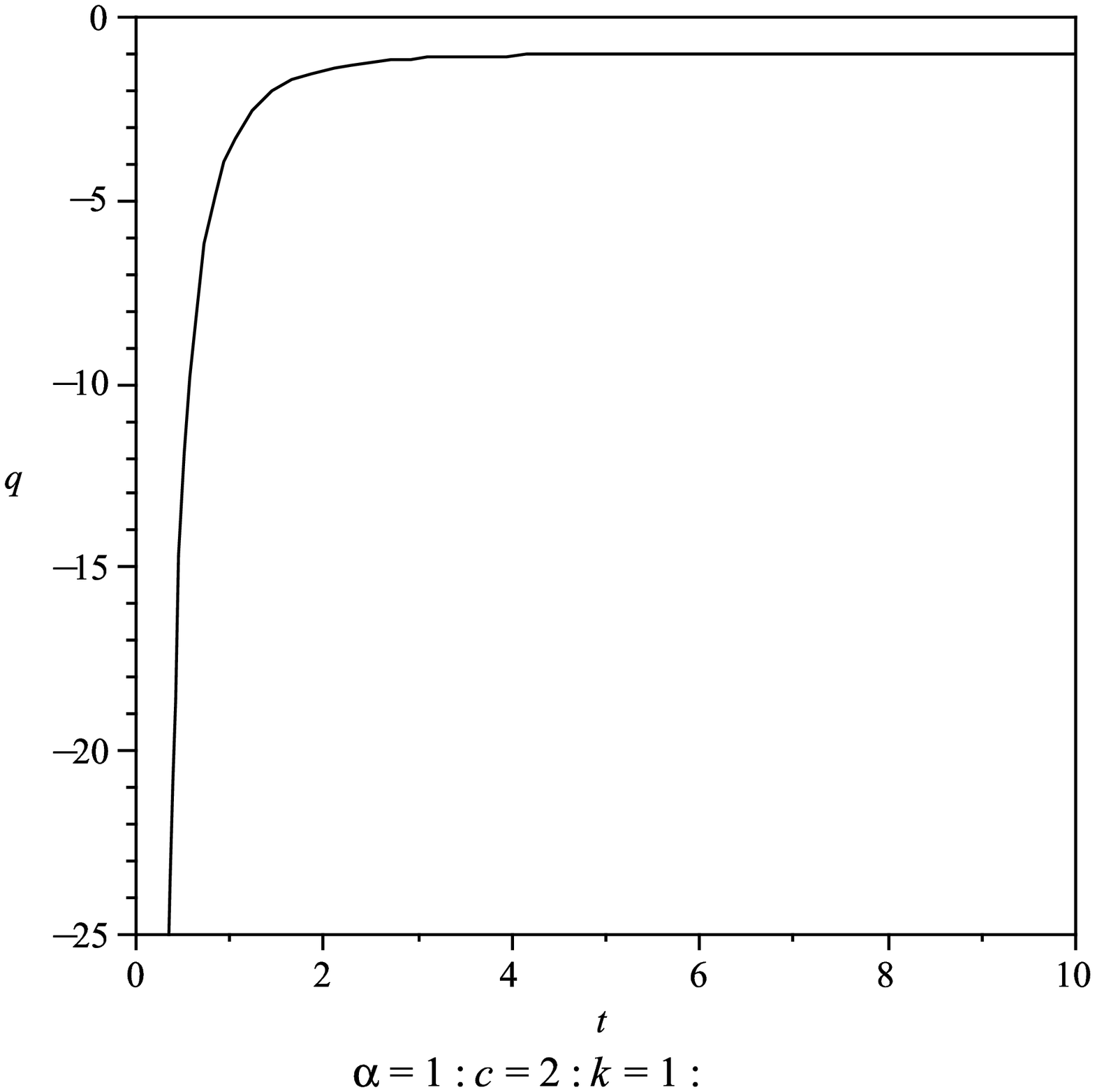}\hspace{0.5cm}\includegraphics[scale=0.25]{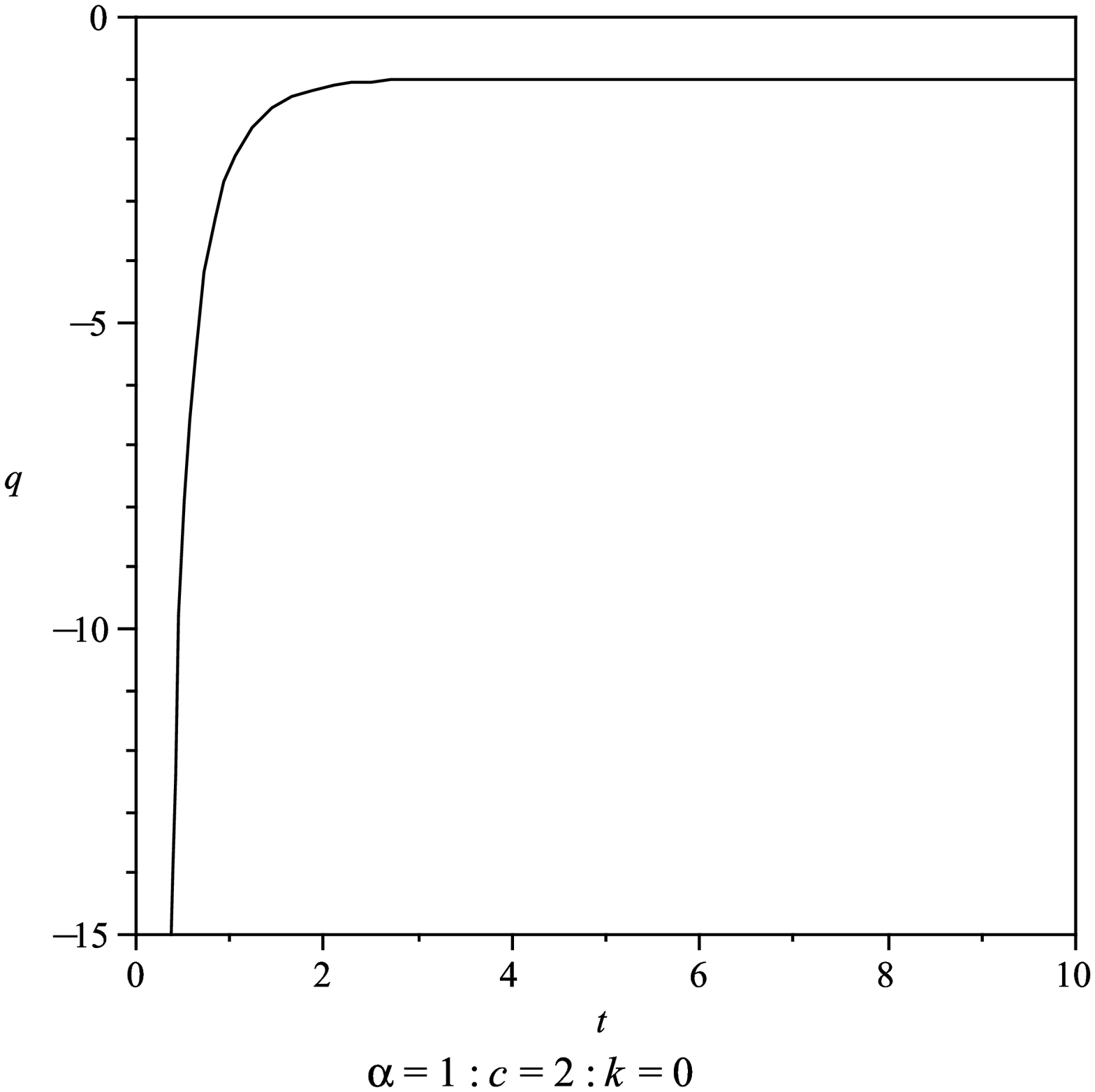}\hspace{0.5cm}
\includegraphics[scale=0.25]{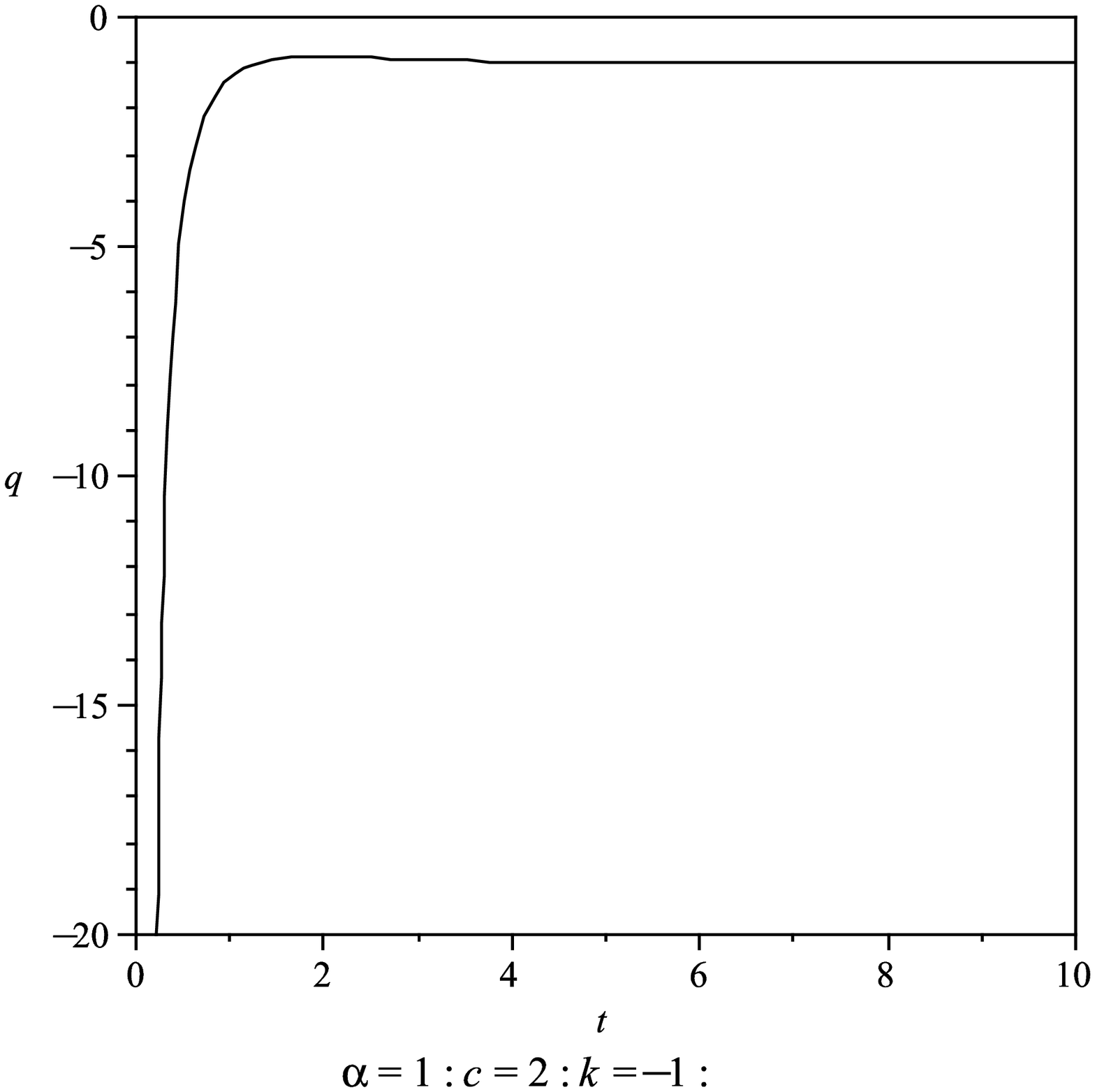}\\
\hspace{-1.5cm}\textbf{Figure 5:} \,Graphs of the acceleration
parameter. \\
\end{tabular*}\\\\\\
By using equations (16) and (18) one can obtain the equation of
state as a following,
\begin{equation}
\omega=-\frac{1}{3}~\frac{3\alpha^2 c^2 \cosh^2(\alpha t)+6k\alpha c
\cosh(\alpha t)+2k^2+\alpha^2 c^2}{\alpha^2 c^2 \cosh^2(\alpha
t)+2k\alpha c \cosh(\alpha t)+2k^2-\alpha^2 c^2}.
\end{equation}\\
We see that in fig. (6) the equation of state for $k=0$ and $k=1$
start to increase and asymptotically approach to negative constant.
But in case of $k=-1$ the equation of state also increases from
positive
value and we see some singularity.\\
\begin{tabular*}{2cm}{cc}
\hspace{0.25cm}\includegraphics[scale=0.25]{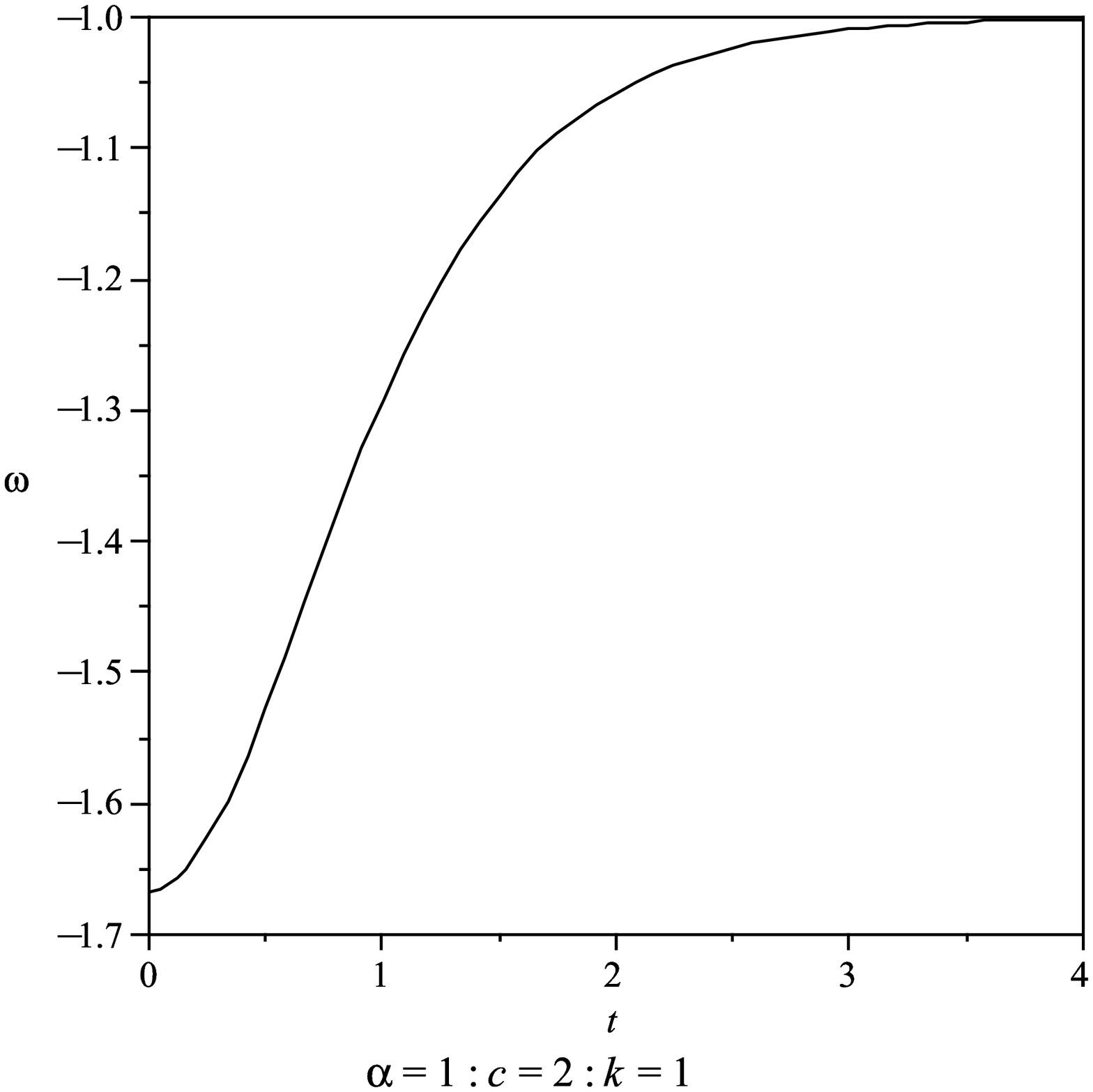}\hspace{0.5cm}\includegraphics[scale=0.25]{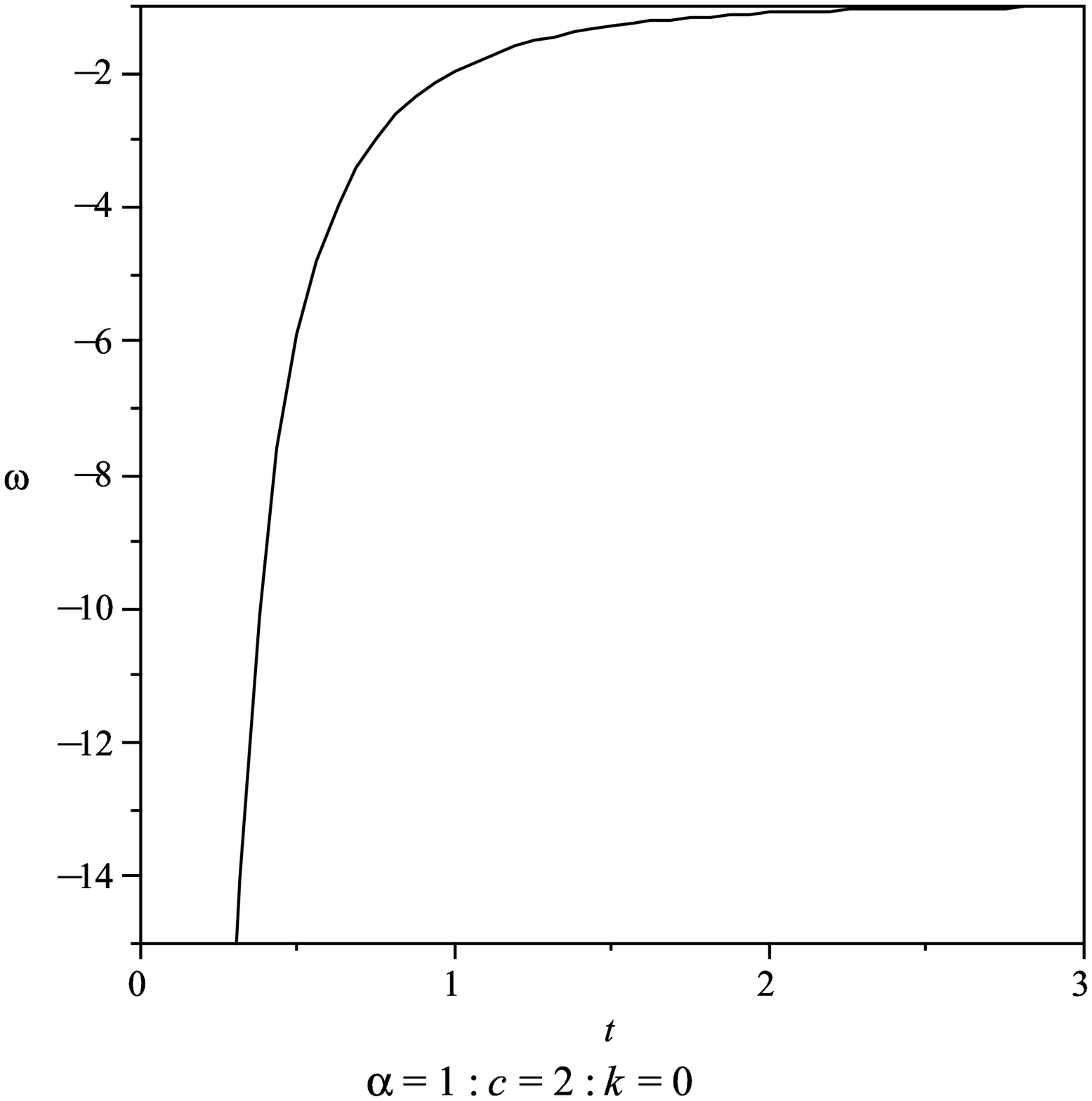}\hspace{0.5cm}
\includegraphics[scale=0.25]{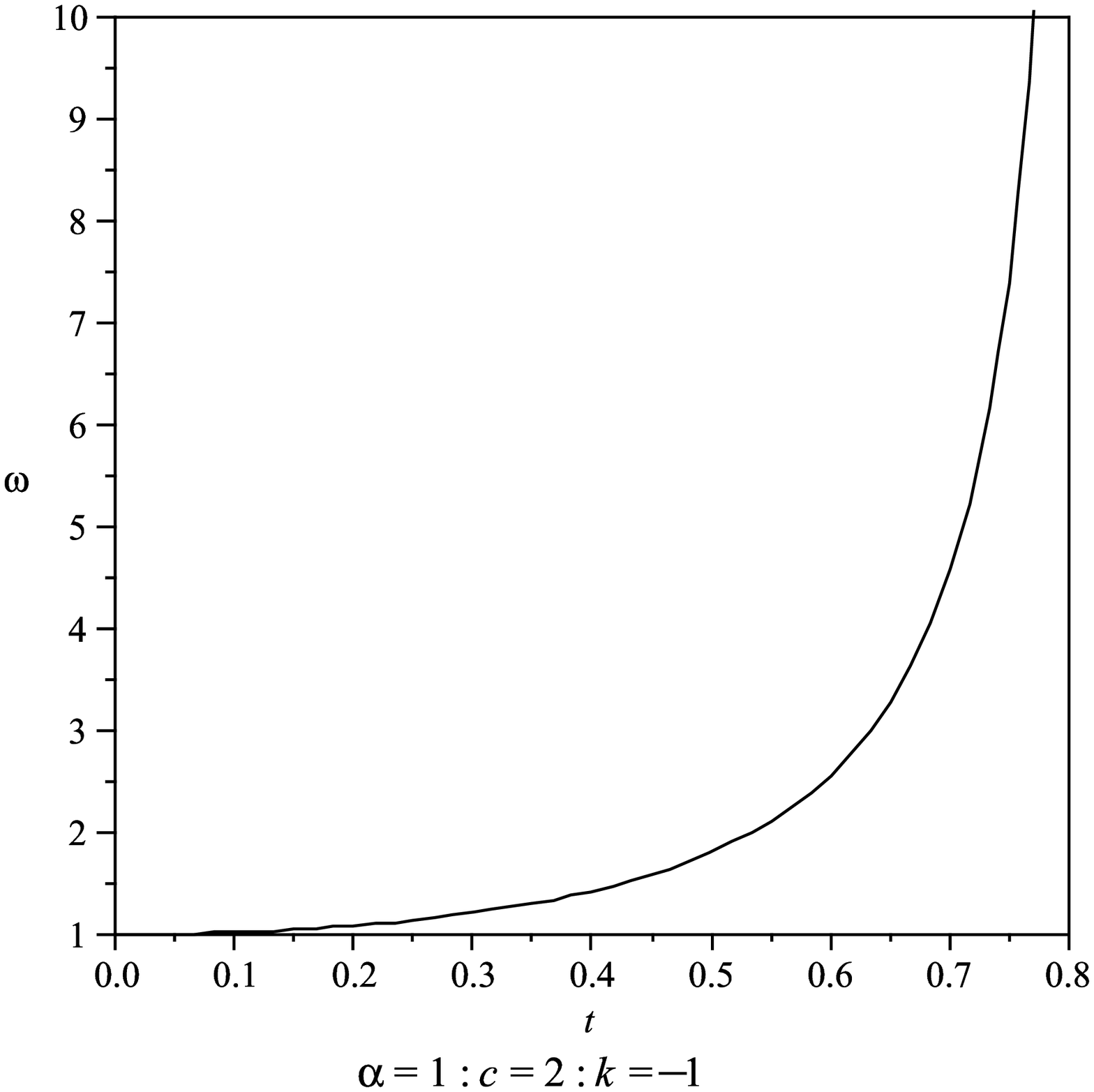}\\
\hspace{-1.5cm}\textbf{Figure 6:} \,Graphs of the equation of state. \\
\end{tabular*}\\\\\\
The corresponding potential for the tachyon field in terms of time
can be obtain by equations (12) and (18),
\begin{equation}
V=\frac{\sqrt{3}~\alpha^2}{8}~~\sqrt{4-\frac{\alpha^2 c^2\left(
\alpha c \cosh^2(\alpha t)+2k \cosh(\alpha t)+\alpha
c\right)^2}{(k+\alpha c \cosh(\alpha t))^4}}.
\end{equation}\\

\begin{tabular*}{2cm}{cc}
\hspace{0.25cm}\includegraphics[scale=0.25]{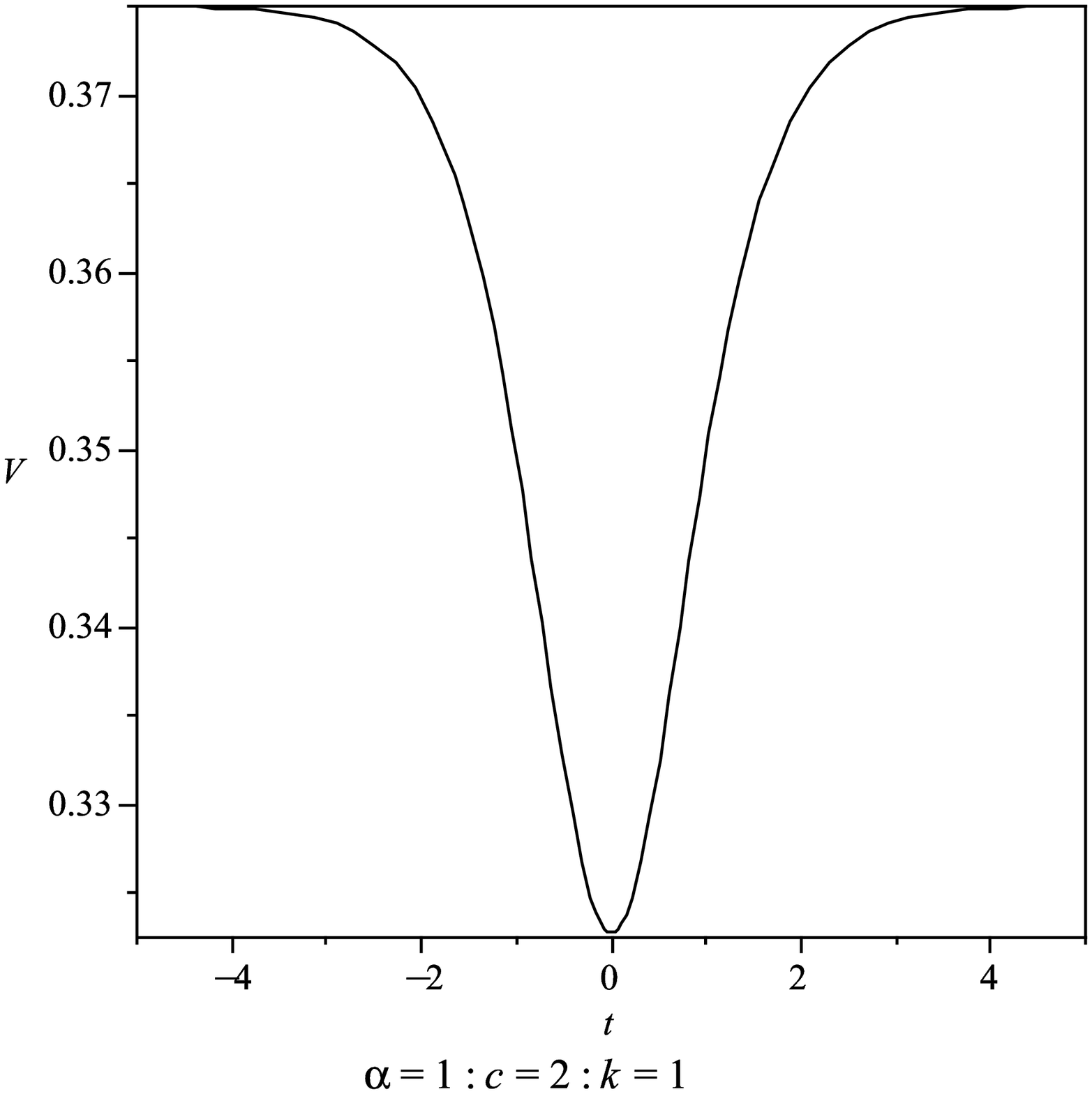}\hspace{0.5cm}\includegraphics[scale=0.25]{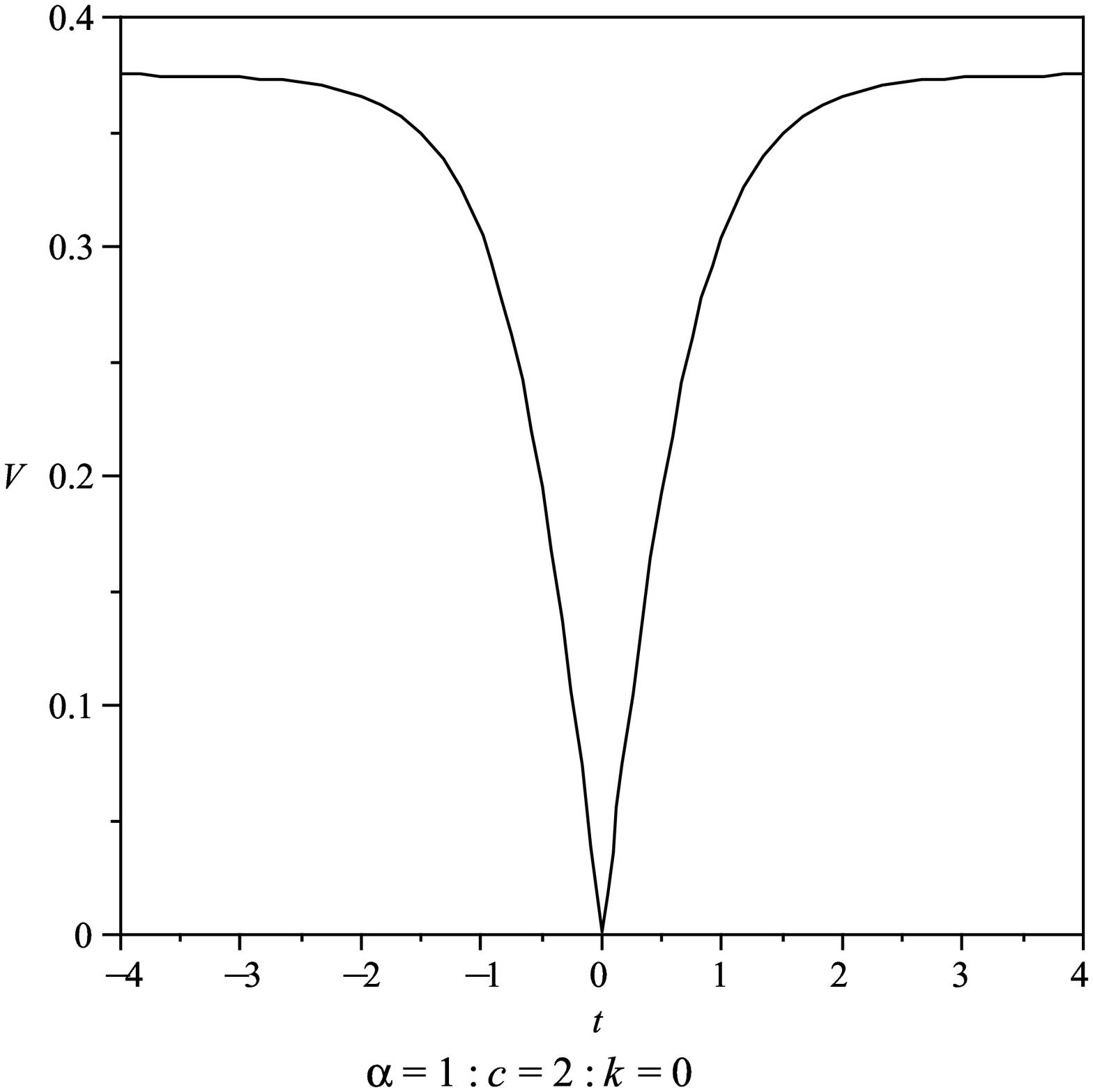}\hspace{0.5cm}
\includegraphics[scale=0.25]{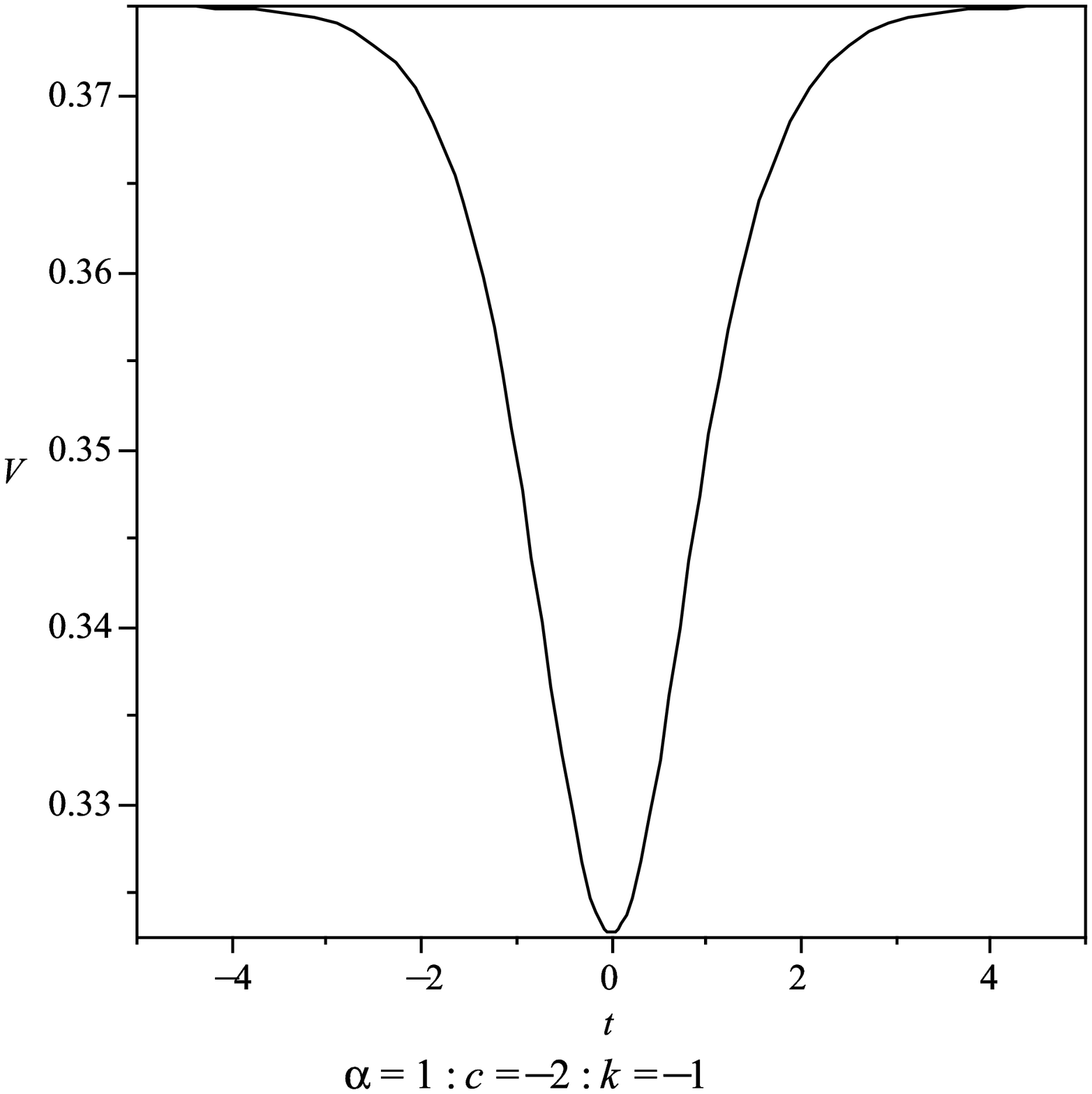}\\
\hspace{-1.5cm}\textbf{Figure 7:} \,Graphs of the potential. \\
\end{tabular*}\\\\\\
In order to obtain the tachyon field we substitute equation (17) in
(13) and draw the following $\dot{T}$ in terms of time, in that case
this graph will be equivalent to $\beta sech(\alpha t)$.
\begin{equation}
\dot{T}=\frac{2\sqrt{3}}{3}~\frac{\sqrt{k^2-\alpha^2 c^2}}{\sqrt{
\alpha^2 c^2 \cosh^2(\alpha t)+2k \alpha c\cosh(\alpha
t)+2k^2-\alpha^2 c^2}}.
\end{equation}\\
and
\begin{equation}
\dot{T}=\frac{2\sqrt{3}}{3}~\frac{\sqrt{k^2-\alpha^2 c^2}}{\alpha c}
~sech(\alpha t)=\beta sech(\alpha t).
\end{equation}\\
so the tachyon field obtains as,
\begin{equation}
T=\frac{\beta}{\alpha}~\arctan(\sinh(\alpha t)).
\end{equation}\\
The behavior of this field in terms of time can be form of
kink-like, which is important to tachyon condensation.
\section{Conclusion}
In this letter, we have considered the curve FRW universe driven by
only tachyonic field. We have presented accelerating expansion of
our universe due to tachyonic field. We also found exact solution of
tachyonic field   which is given by equation (27). This field leads
us to obtained the corresponding potential for the tachyon field. In
this paper we assumed that  $\alpha = constant$ so the $R$ curvature
stays constant.  The interesting  problem  here is to find solution
of tachyon field  for general $f(R)$ instead of $R.$ This problem
will be investigated  in the future.


\begin{thebibliography}{17}

\bibitem{P1}
M. Sami, Mod. Phys. Lett. A {\bf 18} 691 (2003); [hep-th/0205146];
E. J. Copeland, M. Sami and S. Tsujikawa, Int. J. Mod. Phys. D {\bf
15}, 11, 1753 (2006); M. R. Setare, Phys. Lett. B {\bf 653}, 116
(2007); M. R. Setare, Phys. Lett. B {\bf 648}, 329 (2007).
\bibitem{P2}
A. Feinstein, Phys. Rev. D {\bf 66} 063511 (2002);[hep-th/0204140]
\bibitem{P3}
A. Sen, JHEP {\bf 0204}, 048 (2002); JHEP {\bf 0207}, 065 (2002);
Mod. Phys. Lett. A {\bf 17} 1797 (2002);[hep-th/0312153].
\bibitem{P4}
M. R. Garousi, Nucl. Phys. B {\bf 584} 284 (2000); Nucl. Phys. B
{\bf 647} 117 (2002); JHEP {\bf 0305} 058 (2003).
\bibitem{P5}
G.W. Gibbons, Phys. Lett. B {\bf 537} 1 (2002); L. Kofman and A.
Linde, JHEP {\bf 0207} 004 (2002); G. N. Felder and L. Kofman, Phys.
Rev. D {\bf 70} (2004).
\bibitem{P6}
T. Padmanabhan, Phys. Rev. D {\bf 66} 021301 (2002).
\bibitem{P7}
L. R. W. Abramo and F. Finelli, Phys. Lett. B {\bf 575} 165 (2003).
\bibitem{P8}
J. M. Aguirregabiria and R. Lazkoz, Phys. Rev. D {\bf 69} 123502
(2004).
\bibitem{P9}
S. Chattopadhyay, U. Debnath and G. Chattopadhyay, arXiv:0712.3107v1
[gr-qc].
\bibitem{P10}
G.W. Gibbons, Class. Quan. Grav. {bf 20}, s321 (2003)
\bibitem{P11}
E. J. Copeland, M. R. Garousi, M. Sami and S. Tsujikawa, Phys. Rev.
D {\bf 71}, 043003 (2005).
\bibitem{P12}
M. Fairbairn and M. H. Tytgat, Phys. Lett. B {\bf 546}, 1 (2002).
\bibitem{P13}
C. G. Callan, C. Lovelace, C. R. Nappi and S. A. Yost, Nucl. Phys. B
{\bf 308}, 221 (1988).
\bibitem{P14}
R. G. Leigh, Mod. Phys. Lett. A{\bf 4}, 2767 (1989).
\bibitem{P15}
A. Abouelsaood, C. G. Callan, C. R. Nappi and S. A. Yost, Nucl.
Phys. B {\bf 280}, 559 (1987).
\end{thebibliography}
\end{document}